# A Solid-State Dielectric Elastomer Switch for Soft Logic

Nixon Chau[1], Geoffrey Slipher[2,a], Benjamin M. O'Brien[3], Randy A. Mrozek[2], Iain A. Anderson[1, 3, 4]

[1]Biomimetics Laboratory, Auckland Bioengineering Institute, The University of Auckland, Level 6, 70 Symonds Street, Auckland 1010, NZ
[2]U.S Army Research Laboratory, 2800 Powder Mill Road, Adelphi, MD, 20783, USA
[3]StretchSense Ltd, 27 Walls Rd, Penrose, Auckland 1061, NZ
[4]Department of Engineering Science, School of Engineering, The University of Auckland, Level 3, 70 Symonds Street, Auckland 1010, NZ
[a] Author to whom correspondence should be sent (geoffrey.a.slipher.civ@mail.mil)

**Abstract**

In this paper we describe a stretchable solid-state electronic switching material that operates at high voltage potentials, as well as a switch material benchmarking technique that utilizes a modular dielectric elastomer (artificial muscle) ring oscillator. The solid-state switching material was integrated into our oscillator, which self-started after 16s and performed 5 oscillations at a frequency of 1.05Hz with 3.25kV DC input. Our materials-by-design approach for the nickel filled polydimethysiloxane (Ni-PDMS) based switch has resulted in significant improvements over previous carbon grease-based switches in four key areas, namely sharpness of switching behavior upon applied stretch, magnitude of electrical resistance change, ease of manufacture, and production rate. Switch lifetime was demonstrated to be in the range of tens to hundreds of cycles with the current process. An interesting and potentially useful strain-based switching hysteresis behavior is also presented.

Our muscles and nervous system allow us to control our bodies as well as react to stimuli from the environment. The nerve-muscle association involves a close coupling of mechanical and electrical phenomena within a fully integrated and fully soft and pliable mechanism. For soft robotic actuators there has, to-date, been a requirement for a separate electronic controller that performs the control function. In practice, these controllers are rigid structures that are detached from the contours of the robot or device. Thus the current state of the art in soft actuation systems cannot truly be considered soft, are comprised of discrete components, and do not emulate the muscle-nerve association found in biological systems.

Our goal is to produce electronic control that is soft and pliable, and to integrate it within a soft actuator like the natural example. In this paper we show how this goal can be achieved using dielectric elastomer (DE) artificial muscle actuators in combination with a solid-state stretchable electronic switching material. DE can emulate natural muscle operation. Examples include walking[1] and multi-segmented worm-like robots[2], self-configuring minimum-energy micro-grippers[3], tunable soft lenses[4], synthetic jet actuators[5], and active skin for a fish-like blimp[6]. They are also multi-functional[7] and can be used for actuation, sensing, and energy harvesting. But until recently the operation of high voltage DE switching was performed exclusively by conventional hard electronics[1,8].

It is now possible to control the flow of charge on or off the DE device using a piezoresistive Dielectric Elastomer Switch (DES)[8], a soft high voltage resistor that changes resistance by at least three orders of magnitude when stretched. This approach has been demonstrated for automatic charge control on a self-commutating motor[9], a soft energy harvester[10], oscillators[8,11] and an artificial muscle computer[12]. To-date, DES have been made by patterning a conducting carbon grease onto the surface of the DE. When the Grease Switch (GS) is deformed, there is a geometric rearrangement of the conducting filler. At a specif-ic strain, the switch can transition to either a conductive ($10^7$ Ω) or an insulating state ($10^{10}$ Ω). This basic physical behavior has previously been utilized to control movement of electric charge onto or off of the DE (as in the above listed applications), actuating or relaxing it, and thus comprising a basic logic unit that responds to a combination of DC voltage potential and applied stretch.

These basic logic units were previously incorporated into a higher order DE artificial muscle ring oscillator that produced over 1000 oscillations[11]. Implications for this are significant as it forms the foundation for a truly soft robotic device that, supplied with DC voltage, can autonomously produce a lifelike vibratory motion.

However, DES controlled devices are currently far from practical realization. The GS is affected by lifetime and repeatability issues mainly associated with the large electric fields that cause progressive deterioration through vaporization of the grease from arcing. The GS also suffers from a number of other key disadvantages including difficulty in manufacturing, manufacturing repeatability, manufacturing yield, and functional reliability. In addition, the GS has a gradual, not sharp, electronic switching behavior.

In this paper, we introduce a DES material to address some of these issues. We also describe a modular artificial muscle ring oscillator, which allows for rapid benchmarking of switching materials at a more realistically complex system level than is typically used for DE characterization[13]. In the ring oscillator system the DE is mechanically triggering its own charge state transitions by stretching/unstretching the solid-state electronic switch.

The switching material is a polydimethylsiloxane (PDMS) based solid-state high voltage switch using nickel (Ni) powder as a conductive filler. An electric current passing capability occurs when the amount of conductive filler (carbon and silver are other common examples) seeded within an insulating matrix (dielectric) exceeds the percolation threshold[14,15]. The nickel particles are dispersed in the PDMS in such a way as to create an approximately two-dimensional conducting plane configuration in which the planar thickness is between one to three times the characteristic dimension of the filler. The material is then cut to a geometry that approximates a one-dimensional conducting trace collinear with the direction of applied deformation. The restriction of the conductor to approximately one dimension yields a sharp switching behavior when the switch is stretched. The low aspect ratio, approximately spherical, conducting 40 µm diameter nickel particles are distributed in concentrations such that the initial state of the material is just on the conducting side of the percolation threshold at between 1 – 5 kV potential. Upon an applied deformation the PDMS elastomer matrix between the particles allows the spacing of the conductive filler to increase in the direction of the load until the percolation threshold is crossed and the switch becomes electrically insulating, switching off. When the load is removed, the elastomeric material returns to its nominal geometry and distribution, thus re-crossing the percolation threshold, and reversing the switching behavior.

A high viscosity reactive PDMS precursor (100,000 centiStokes) was mixed in equal parts with a low viscosity non-reactive PDMS precursor (10 cSt) to reduce the viscosity of the mixture to 6,500 cSt and allow the filler to settle out slowly over the three-day cure time[16]. Nickel concentrations were evaluated in a range from 20 % to 40 % of PDMS by weight. This resulted in a thin, approximately two-dimensional (100 µm) conducting layer on the bottom and a much thicker (900 – 1500 µm) depletion layer on the top. The resulting cured material was therefore conductive on only one surface. The material was then cut into the dog-bone shapes as shown in Figure 1 using a CNC laser cutter. We ended up selecting the 25% Ni content switches for the work that is presented in this paper. We did so because Ni contents higher than 25% have lower voltage ranges at which they exhibit a switching behavior, and we needed them to switch in the range of 3kV in order to effectively interact with the DEA components of the oscillator. Lower concentrations than 25% require higher voltage potentials to be applied in order to exhibit the switching behavior. The limits of this trend were not explored.

Seven samples of the Ni-PDMS switch (Table 1; Figure 1) were clamped and cyclically stretched up to 77% strain using a servotube (Copley Motion STB2504) at 0.5Hz and 1Hz. Resistance was measured at a voltage of 2 kV. The highest number of cycles to failure was 380. The lowest cycles to failure was 5. Average lifetime (cycles) was 99.6. Failure was manifested by a switch persisting in the 'on' (low resistance) state. The number of samples and tests conducted needs to be extended in order to reliably compute statistics on repeatability, average switch lifetime, and activation strain.

| No | Voltage (kV) | Max Cycles (n) | Freq (Hz) | Average Activation/Deactivation strain |
|---|---|---|---|---|
| 1 | 2 | 45 | 1 | 0.35/0.15 |
| 2 | 2 | 86 | 1 | 0.2/0.09 |
| 3 | 2 | 380 | 0.5 | 0.45/0.31 |
| 4 | 2 | 5 | 0.5 | 0.37/0.17 |
| 5 | 2 | 32 | 0.5 | 0.42/0.15 |
| 6 | 2 | 100 | 0.5 | 0.6/0.48 |
| 7 | 2 | 49 | 0.5 | 0.48/0.12 |

Table 1: Data for Ni-PDMS samples.

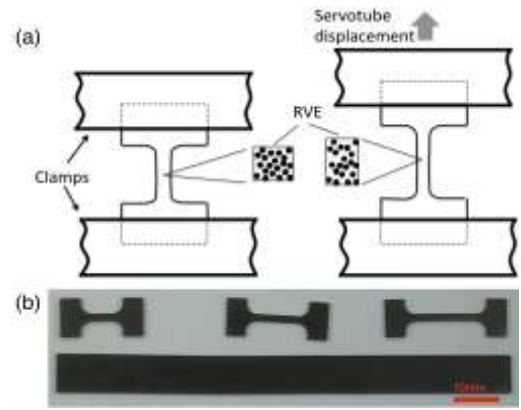

Figure 1: a) Switch material strained by a servotube. The representative volume element (RVE) depicts stiff Ni particles remaining undeformed, but with increased inter-particle spacing in the strained sample. b) Different Form Factors of the Ni-PDMS Switch. These were cut using a laser cutter.

The Ni-PDMS composite material exhibited a switching behavior similar to an ideal switch where the switching is nearly instantaneous in the strain domain (Figure 2). The resistance saw a vertical jump from $10^6$ $\Omega$ to approximately $10^{11}$ $\Omega$ at approximately 30 % tensile strain, which corresponds to the conductive and insulating states of the material.

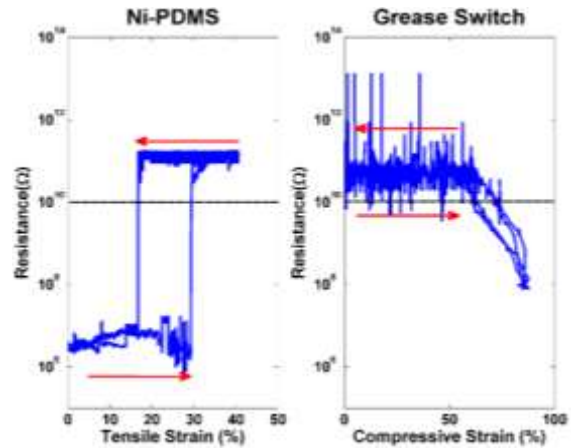

Figure 2: Switching Performance of both dielectric elastomer switches. The Ni-PDMS DES switches from low resistance to high resistance with increasing tensile strain. The grease switch, patterned on a stretched elastomer membrane (VHB 4905 from 3M ™) switches from high to low resistance on relaxation of the taut membrane; on increasing compressive strain.

The Ni-PDMS switch material provides a two order-of-magnitude improvement in resistance change over the GS, and is completely soft (approximately 200kPa bulk modulus) and stretchable (>50% strain to break). The strain magnitude at which switching occurs is con-trolled by how close the initial particle density is to the percolation threshold. The initial conducting or insulating state can be controlled such that it is either insulating (by applying a sufficient pre-strain to cross the percolation threshold), or conducting (by applying either modest or no pre-strain).

The Ni-PDMS production process also provides significant improvements in ease of manufacture and rate of production over the GS. In our extensive experience with producing devices based upon the GS, the typical production time per switch is 5 minutes, including all preparation work, and the typical functional switch yield is 75%. This we contrast with the production of the Ni-PDMS based switches which involve a total of about two hours preparation time combined with a 72 hour cure resulting in hundreds or thousands of switches depending on batch size and switch geometry. Yield for the Ni-PDMS process is close to 100% for exhibiting reliable switching behavior and a current average switch lifetime of approximately 100 cycles (see Table 1).

We have produced and demonstrated a ring oscillator for benchmarking switching materials (Figures 3, 4)[17]. The design of the artificial muscle ring oscillator was inspired by a previous oscillator[11]. In this modular version, two DE actuators (DEAs) that pull the single Ni-PDMS DES opposite in tension were required in order to generate sufficient strain for switch activation (Figure 4). A modular design was also adopted to allow for rapid material benchmarking and easy interchange of different switching materials.

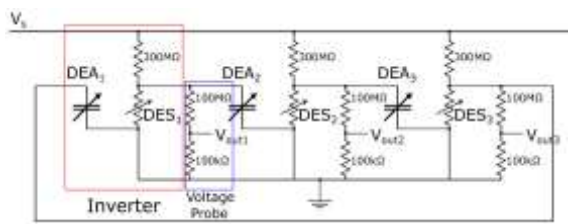

Figure 3: Circuit schematic of an artificial muscle ring oscillator. Note that DEA and DES are depicted as variable capacitors and variable resistors respectively.

The DEA were constructed from 3M VHB (Very High Bond) 4905, an acrylic-based elastomeric film that was pre-stretched equibiaxially to a stretch ratio of 3.5 resulting in a final approximate thickness of 40 μm, and secured to a rigid acrylic frame. Nyogel 756G carbon grease was applied onto both sides of the film to form the electrodes. These elements undergo planar expansions as a voltage is applied across the electrodes associated with the electrostatic Maxwell pressure[18].

Rigid acrylic bars connected one end of the switch to the DEA and the linear stage on the other end, which was fixed in place (see Figure 4). A 300 MΩ resistor was placed in series with the switch and this whole unit (with two DEA attached together), comprises an inverter, as shown in Figures 3 and 4.

The 300 MΩ resistor value was selected to be in between the two switch states ($10^6$ Ω for low and $10^{10}$ Ω for high) such that the output voltage across the switch generated the largest amplitude. Therefore, as the switch resistance alternated, the output voltage fluctuated closely between ground and the supply voltage. Three of these inverter units were required in order to create the complete three-stage ring oscillator represented in the Figure 3 schematic.

The oscillator system was evaluated with an initial supply voltage of 3 kV. The supply voltage was progressively increased by increments of 250 V until oscillation initiated, or the formation of wrinkling around the edges of the VHB membrane was present. Wrinkling signaled the onset of DEA failure.

Demonstration of Ni-PDMS DES operation can be seen in Figure 5. In this example the ring oscillator self-started after 16 seconds at 3.25 kV, and oscillated for 5 cycles with a frequency of approximately 1.05 Hz before failure occurred.

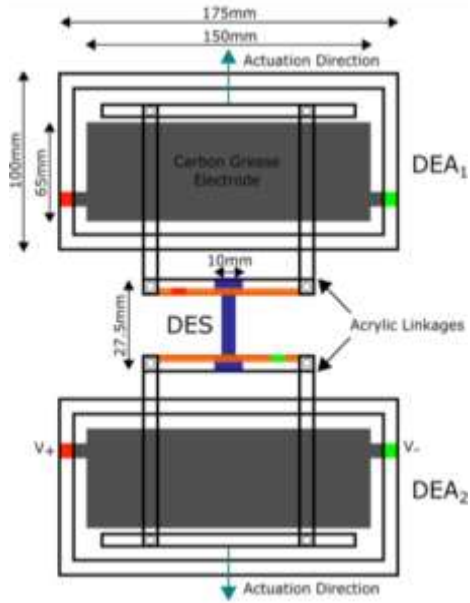

Figure 4: Graphical representation of an inverter unit consisting of 2 DEAs pulling the Ni-PDMS DES in tension.

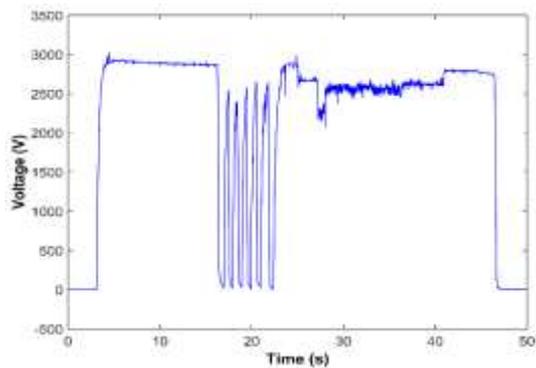

Figure 5: Output voltage of modular ring oscillator using ARL's Ni-PDMS switching material with 3.25 kV DC input. Note that when the switch material failed, it remained in a conducting state (low resistance) starting at 23 seconds until the end of the experiment when the DC input was switched off.

The introduction and demonstration of the solid-state Ni-PDMS switching material has presented some immediate advantages over the grease-based switches. The solid-state elastomer switching material provides a two order-of-magnitude increase in resistance change over the GS, which allows for greater circuit flexibility as well as more defined states. Furthermore, the sharper strain-induced switching behavior will aid in generating faster oscillation frequencies. The hysteretic switching profile in which the Ni-PDMS switch material transitions between insulating/conducting states at different strain magnitudes (see Figure 2) could be exploited to form a mechano-sensitive hysteretic comparator circuit (e.g. Schmitt trigger).

Despite the improvements over the GS, some significant challenges remain unresolved with the current Ni-PDMS material. This switch material is prone to dielectric breakdown, whereby it remains stuck in a conducting state after a relatively small O(100) number of stretching cycles. A seamless, fully soft integration of the switch with the DE artificial muscle also remains undemonstrated. These challenges may be mitigated through further innovations to the switch material processing, such as manipulation of the interfacial bond strength between the conductive filler and the elastomer matrix, and integration of the full electromechanical system into a single manufacturing process.

Although our initial attempt at a ring oscillator with the Ni-PDMS switching material only managed to produce 5 cycles, it has demonstrated the feasibility of the solid-state material to function in a complex integrated system. We attribute the disparity between individual switch lifetimes in the hundreds of cycles and the much shorter cycle lifetime of the assembled ring oscillator system to the reduced reliability of multiple dielectric elastomer elements (switches with actuators) operating within the same system.

Possible avenues for creating a more robust benchmarking oscillator system include using different conductive materials for the DEA electrodes to minimize actuator failure due to dielectric breakdown and arcing. Examples of candidate electrode materials are self-clearing carbon nanotubes[19] or graphene. DEA geometry optimization to generate consistent strain of the right magnitude for the DES will also help to improve reliability by minimizing strain excursions that can potentially damage the switch material. Utilization of conductive fillers for the switch material with smaller characteristic dimension, such as carbon nanotubes, graphene, or colloidal gold/silver are worth exploring as they would have the benefit of enabling reduced switch size. The artificial muscle ring oscillator has some promising applications: a device which can be used as a switch that responds to both electrical and mechanical inputs; as a clock signal for peristaltic pumps, simple periodic motions, and motors; or as a central pattern generator for motion control in soft robotic systems. Effort must be made in improving lifetime, which currently plagues both types of DES materials (grease and Ni-PDMS).

The current modular ring oscillator setup serves as a benchmarking platform whereby other potential DES materials could be rapidly tested and compared to current materials, thereby accelerating the development of improved DES.

In this paper we have demonstrated a modular artificial muscle ring oscillator using a purposefully designed solid-state Ni-PDMS elastomer DES. We have demonstrated the efficacy of a focused materials-by-design approach for yielding significant performance improvements over the state-of-the-art carbon grease-based switches. We have shown a two order-of-magnitude increase in resistance change upon applied stretch and have dramatically improved the sharpness of the switching behavior to be nearly instantaneous in the strain domain. We have simultaneously improved the production rate and yield. We have also demonstrated evidence of a potential mechanosensitive hysteretic comparator circuit based on strain-hysteretic behavior in the Ni-PDMS switch material.

Soft devices could one day be seamlessly integrated with us in much the same way that nerves and muscles form locally emergent closed-loop control systems distributed throughout our bodies. A reliable switching material that seamlessly translates information between the electrical and mechanical domains is a necessary development along the path to achieving such a vision.

**Acknowledgements**

N.C. and I.A. acknowledge the U.S. Army International Technology Center, Pacific (USAITC-Pac) for providing funding for the work performed at the University of Auckland under grant number FA5209-14-P-0016.

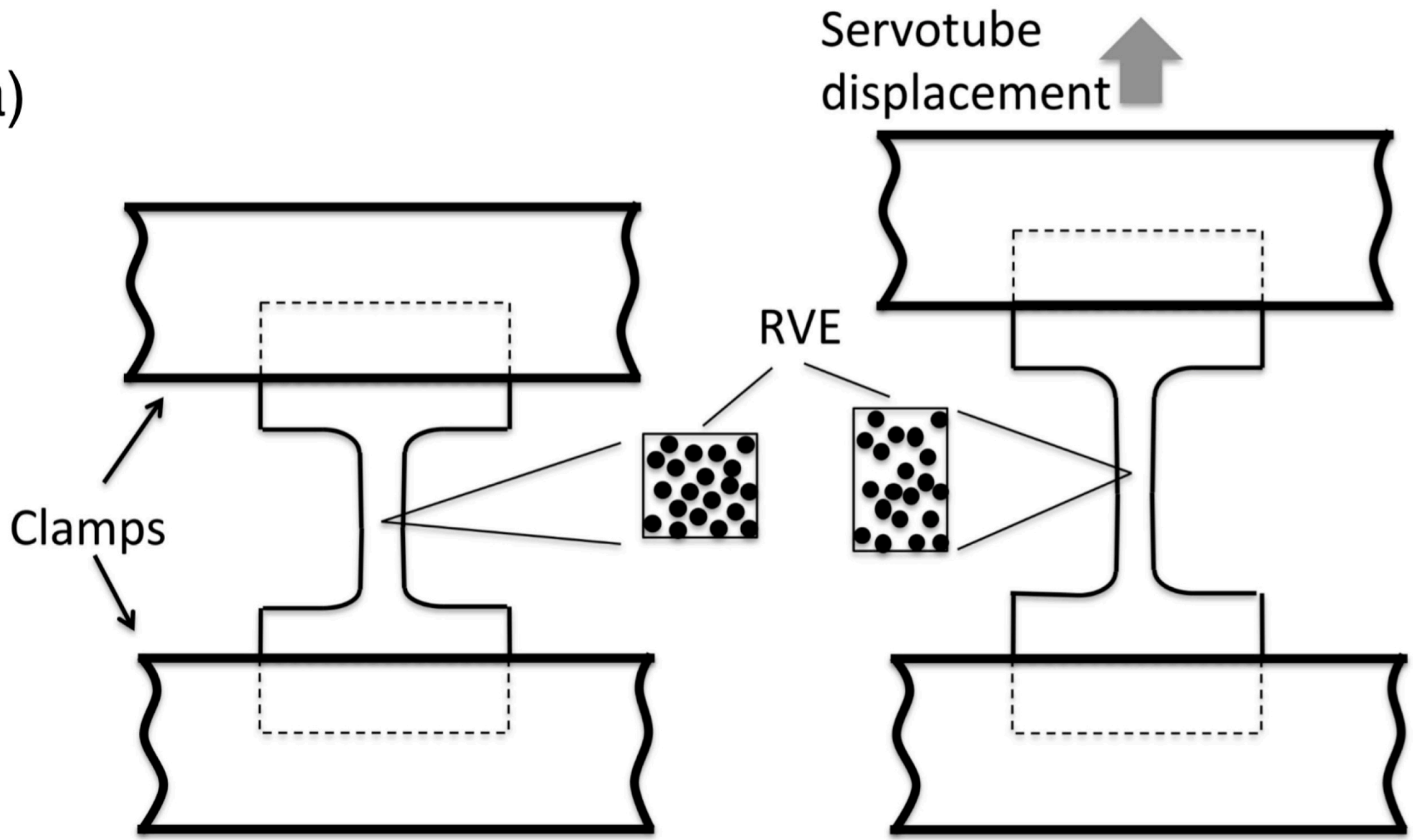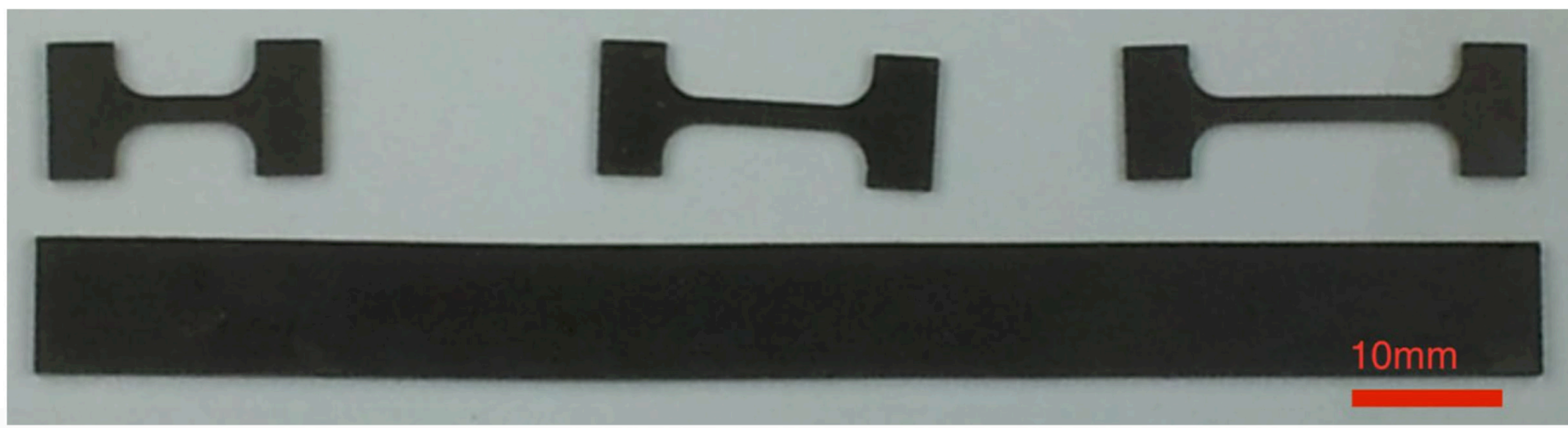

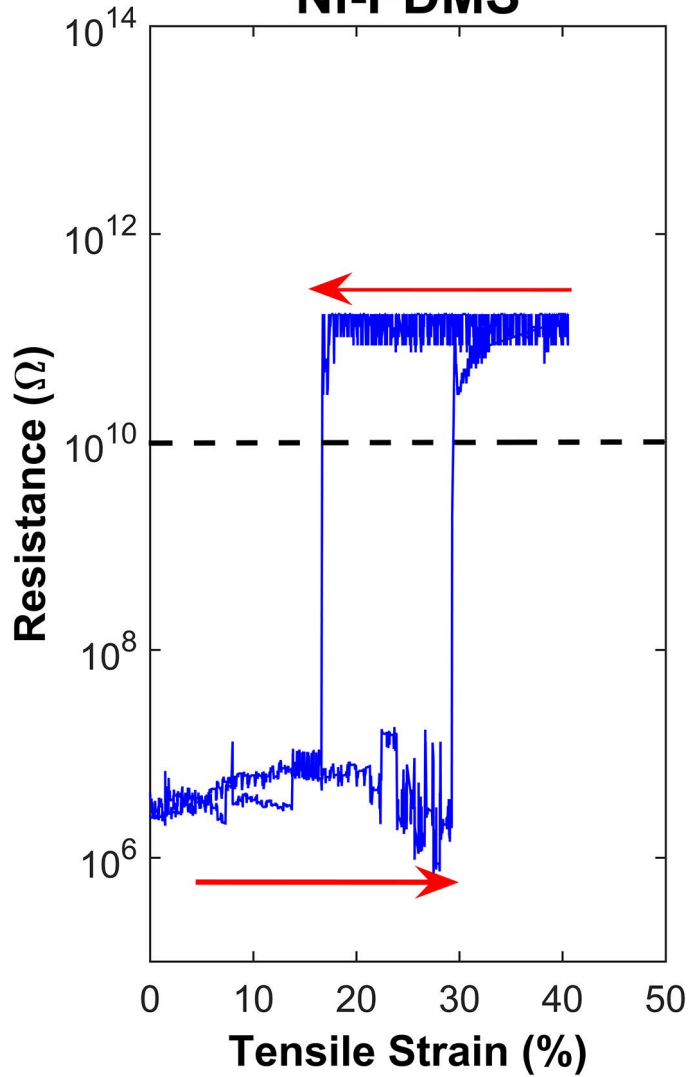 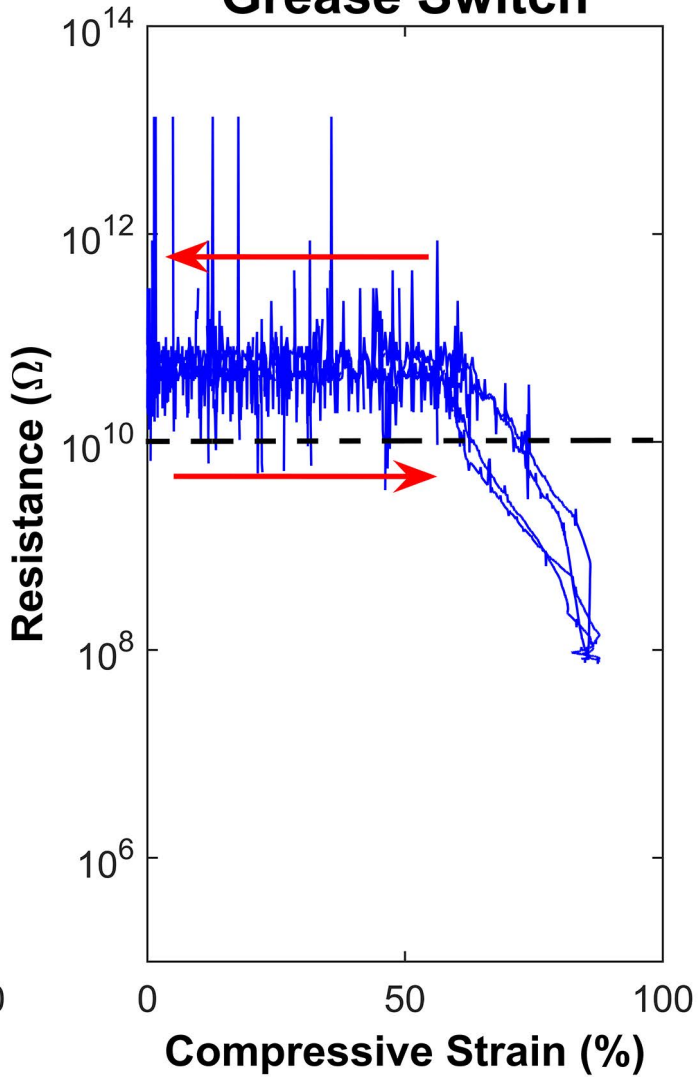

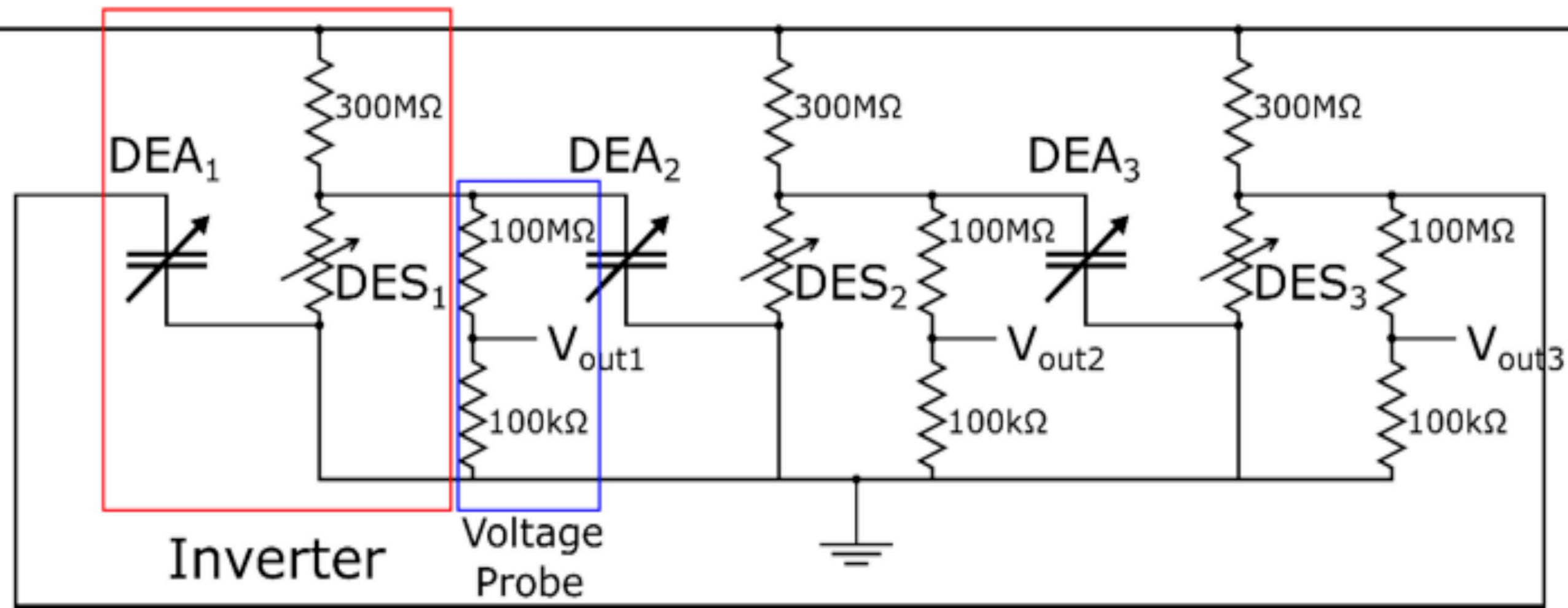

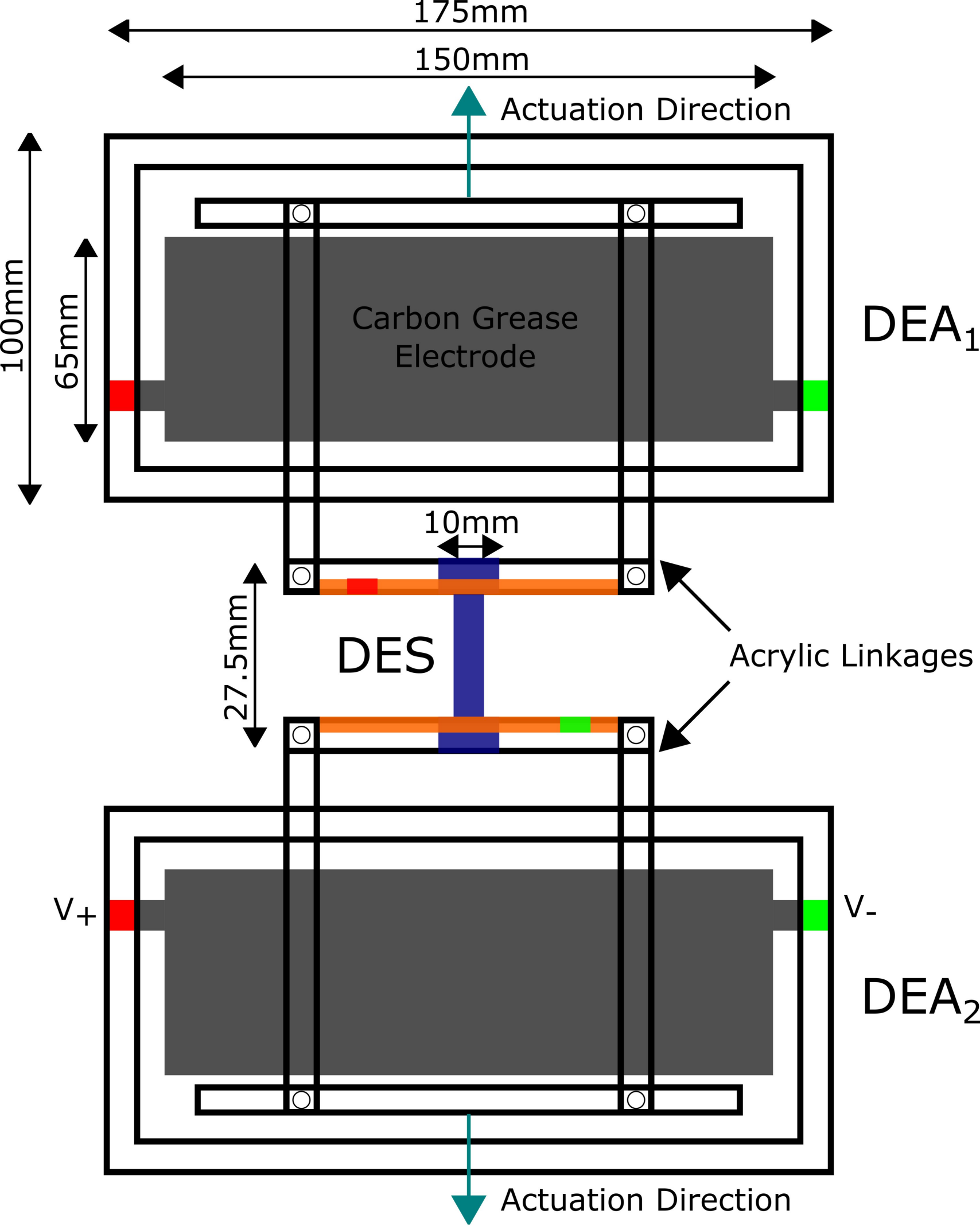

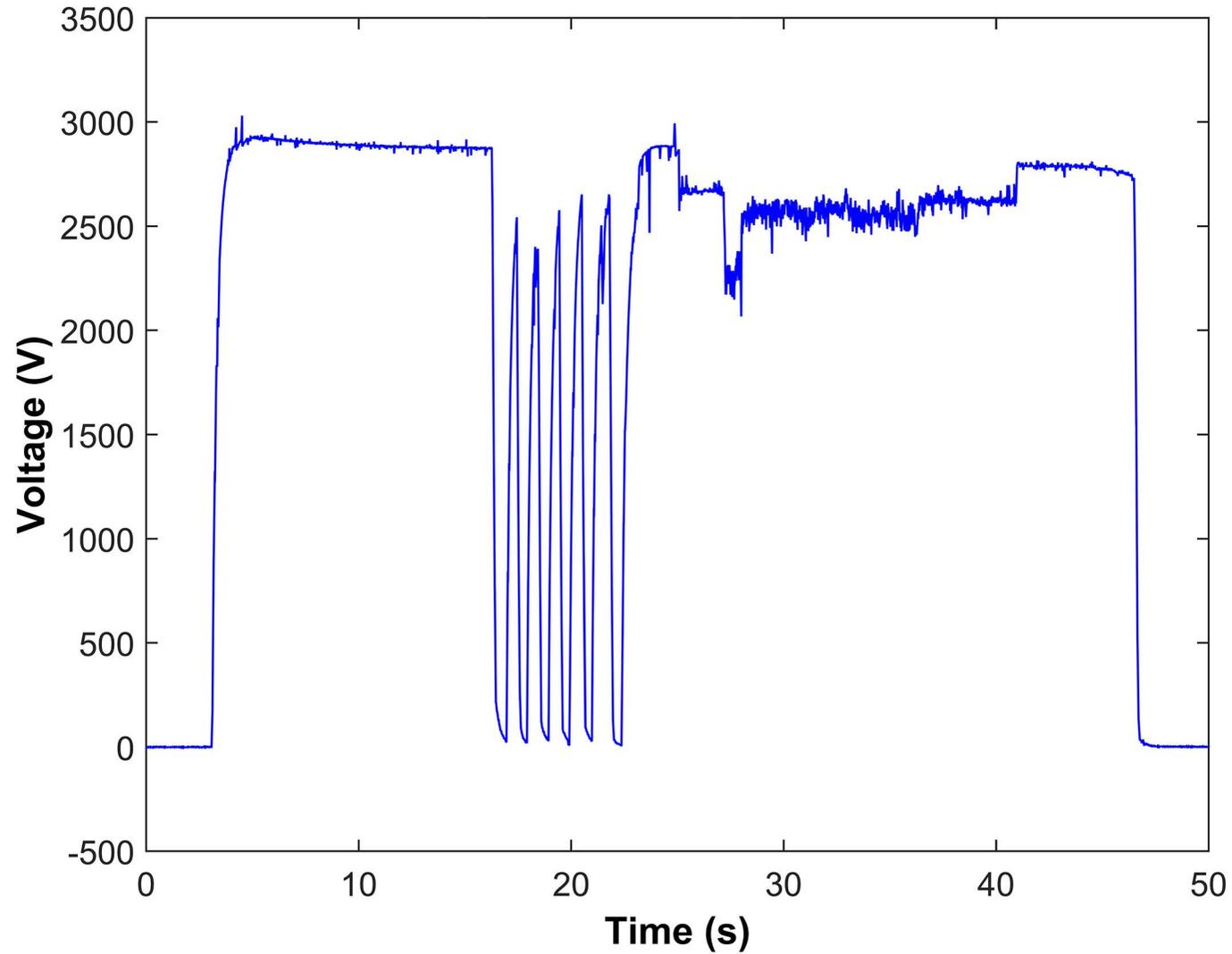